\documentclass[letterpaper]{JHEP3}

\usepackage{graphicx,epsfig,citesort}
\usepackage{latexsym}

\title{Domain wall solutions with Abelian gauge fields}

\author{
J.S. Rozowsky \\
Physics Department, Syracuse University, Syracuse, NY 13244-1130, U.S.A.\\
\email{rozowsky@phy.syr.edu}}

\author{
R.R. Volkas\\
School of Physics, Research Centre for High Energy Physics,
The University of Melbourne, Victoria 3010, Australia.\\
\email{r.volkas@physics.unimelb.edu.au}}

\author{
K.C. Wali \\
Physics Department, Syracuse University, Syracuse, NY 13244-1130, U.S.A.\\
\email{wali@phy.syr.edu}}

\abstract{
We study kink (domain wall) solutions in a model consisting of two 
complex scalar fields coupled to two independent Abelian gauge fields 
in a Lagrangian that has $U(1)\times U(1)$ gauge plus $\mathbb{Z}_2$ 
discrete symmetry. 
We find consistent solutions such that while the $U(1)$ symmetries 
of the fields are preserved while in their respective vacua, they are 
broken on the domain wall. The gauge field solutions show that 
the domain wall is sandwiched between domains with 
constant magnetic fields.}

\preprint{SU-4252-782}

\begin{document}

\section{Introduction}

Over the last thirty years or so,
the study of solitonic solutions to classical field theories has yielded
many interesting results of wide
relevance to particle physics, cosmology and condensed matter physics.
The more recent fascination with brane-world models of particle
physics and cosmology has added new motivation for these kinds of
investigations. In this paper we will study a simple model of
two complex scalar or Higgs fields $\phi_1$ and $\phi_2$ coupling to two 
different $U(1)$ gauge fields $A_{1\mu}$ and $A_{2\mu}$,
with the added feature of an exact discrete $\mathbb{ Z}_2$ symmetry 
under the interchange $1 \leftrightarrow 2$. We will derive solutions
to the coupled classical field equations that exhibit a kink or domain 
wall form for the scalar fields. The nature of the gauge field
configurations self-consistently coupled to the Higgs kinks
will be our primary object of study. A similar model, without the
discrete exchange symmetry was studied some-time ago by 
Witten~\cite{Witten} in the context of a superconducting 
string solution. The model was investigated in more detail by 
MacKenzie~\cite{MacKenzie} to show that while a symmetry is 
preserved in the vacuum, unexpected topological structures 
can arise in the interior of a domain wall. More recently, 
Lemperiere and Shellard~\cite{Lemperiere} have reported 
on the behavior and stability of the superconducting currents 
in Witten's model.

Our own motivation for this rather abstract investigation lies
with the symmetry breaking mechanism proposed in Ref. \cite{clash}
in the context of brane world models and dubbed as the 
``clash of symmetries''.  Briefly, Ref. \cite{clash}
examines a toy model with Higgs fields in three triplet representations
of a {\it global} $SU(3)$ symmetry, where a discrete
permutation symmetry between the triplets is enforced. Omitting
inessential complications, the vacuum
states of the theory spontaneously break $SU(3)$ down to $SU(2)$,
as well as spontaneously breaking the discrete symmetry. Kink
solutions are derived that interpolate between vacua invariant
under {\it differently embedded} $SU(2)$ subgroups.\footnote{Qualitatively
similar solutions, but to a different theory with a different motivation
were discovered by Pogosian and Vachaspati in Ref. \cite{pv}.} 
For instance,
one can have $I$-spin asymptotically preserved 
on one side of a domain wall, 
with $V$-spin on the other. Although the unbroken subgroups
on both sides are isomorphic, the different embeddings within the
parent group cause additional symmetry breakdown at all
non-asymptotic points. This additional symmetry breaking is
the ``clash''. The idea is that some of the symmetry breaking
we see in our universe might be due to such a clash, if our world
is indeed a brane in a higher dimensional space.

This idea is still at the developmental stage; no realistic
brane-world model building using the clash mechanism has yet been
attempted, to our knowledge, though Ref.~\cite{Shin} reports
on some recent progress. In the course of thinking about
the clash of symmetries idea, however, an even simpler model field
theory with $U(1)$ factors and interchange symmetries between
the different sectors naturally presented itself as a
useful theoretical laboratory. The model studied in this
paper arose in exactly this way, though, of course, it is
also entitled to an independent existence as a 
simple-but-not-too-simple vehicle for the study of gauge
fields coupled to domain wall Higgs configurations. From
this perspective, our work is relevant to general
studies of superconducting topological solitons, as in
Refs.~\cite{Morris,Witten,Lemperiere,MacKenzie,Lazarides} 
for example. From the clash of
symmetries perspective, the present exercise begins the
study of the breakdown of {\it local} continuous symmetries.

The rest of this paper is structured as follows: In Sec. 2,
the model and the field equations are presented. The
numerical study of kink solutions to these equations is
then presented in Sec. 3, while Sec. 4 provides a physical
explanation for the solutions. Section 5
contains some concluding remarks.

\section{The Model}

Using the the notation of \cite{clash} we start with
the action for two complex scalar fields $\phi_{1,2}$
coupled to different
$U(1)$ gauge fields $A_{1,2}$. To the overall $U(1) \times U(1)$ gauge
symmetry we add a $\mathbb{ Z}_2$
discrete symmetry which interchanges the scalars, $\phi_1 \leftrightarrow \phi_2$
and the gauge fields, $A_1 \leftrightarrow A_2$. The discrete symmetry
makes the two gauge coupling constants equal in magnitude. 
The Lagrangian is
\begin{eqnarray}
{\cal L} = -{1\over 4}F_1^{\mu\nu} F_{1\mu\nu}
-{1\over 4} F_2^{\mu\nu}F_{2\mu\nu}
+\left(D_1^{\mu}\phi_1\right)^* \left(D_{1\mu}\phi_1\right)
+\left(D_2^{\mu}\phi_2\right)^* \left(D_{2\mu}\phi_2\right) 
-V(\phi_1,\phi_2),
\end{eqnarray}
where 
\begin{eqnarray}
V(\phi_1,\phi_2) = 
\lambda_1\left(\phi_1^*\phi_1+\phi_2^*\phi_2-\upsilon^2\right)^2
+\lambda_2\phi_1^*\phi_1\phi_2^*\phi_2.
\end{eqnarray}
The covariant derivatives in the Lagrangian are given by
\begin{eqnarray}
D_{1\mu} = \partial_\mu-ieA_{1\mu} \qquad, \qquad 
D_{2\mu} = \partial_\mu-ieA_{2\mu}. 
\end{eqnarray}

The Higgs potential admits two vacuum solutions:
\begin{eqnarray}
{\rm Vacuum \: 1:} & \langle\phi_1^*\phi_1\rangle = \upsilon^2 &
\langle\phi_2^*\phi_2\rangle = 0, \\
{\rm Vacuum \: 2:} & \langle\phi_1^*\phi_1\rangle = 0 &
\langle\phi_2^*\phi_2\rangle = \upsilon^2.
\end{eqnarray}
These two vacua are degenerate and are the global minima of 
the potential for the parameter regime
\begin{eqnarray}
\lambda_1\geq 0 \quad {\rm and} \quad \lambda_2\geq 0.
\end{eqnarray}

We would like to construct domain wall solutions by requiring the
scalar Higgs fields to asymptote to 
different respective vacua on either side 
of the wall. We will be interested in the behavior of the
corresponding gauge fields for this kind of Higgs configuration.
The boundary conditions for the scalars are
\begin{eqnarray}
|\phi_1(z)| = \left\{
\begin{array}{lll}
0  & \quad & z\rightarrow-\infty \\
\upsilon & \quad & z\rightarrow\infty 
\end{array}
\right. 
\quad {\rm and} \quad
|\phi_2(z)| = \left\{
\begin{array}{lll}
\upsilon  & \quad & z\rightarrow-\infty \\
0 & \quad & z\rightarrow\infty 
\end{array}
\right. ,
\label{BC1}
\end{eqnarray}
where $z$ is the direction perpendicular to the domain wall.

It is straightforward to compute the equations of motion for the
Higgs fields
\begin{eqnarray}
D_{a\mu}D_a^\mu \phi_a = -{\partial V\over \partial \phi_a^*}
=-2\lambda_1\phi_a\left(\phi_a^*\phi_a+\phi_b^*\phi_b-\upsilon^2\right)
-\lambda_2\phi_a\phi_b^*\phi_b,
\end{eqnarray}
where $a,b$ are either $1,2$ or $2,1$ respectively. 
The equations of motion for the gauge fields are similarly given by
\begin{eqnarray}
\partial_\mu F_a^{\mu\nu} = 2e \,{\rm Im}\left[\phi_a^*
(\partial^\nu-ieA_a^\nu)\phi_a\right].
\end{eqnarray}

Since we are going to be looking for static domain wall 
solutions (i.e. static $1+1$ solitons), we search for solutions
that depend on $z$ but are independent of all the other spatial
coordinates and time $t$. 
In order to simplify our equations we make use of the temporal gauge, 
$A_0=0$. With these choices the equations of motion reduce to
\begin{eqnarray}
A_{1z} &=& {\alpha_1'\over e}, \label{PureGauge}\\
A_{1x,y}'' &=& 2e^2 A_{1x,y} R_1^2, \\
R_1'' &=& e^2(A_{1x}^2+A_{1y}^2)R_1 + 2\lambda_1 R_1(R_1^2+R_2^2-\upsilon^2)
+\lambda_2 R_1 R_2^2,
\end{eqnarray}
where prime denotes a derivative with respect to $z$ and 
$\phi_a \equiv R_a(z) e^{i\alpha_a(z)}$. The corresponding equations
for the fields with subscript $2$ can be obtained simply by exchanging
subscripts $1$ and $2$. 
We see in eqn.~\ref{PureGauge} that the $z$ components of both 
gauge fields are pure gauge and because neither $A_z(z)$ nor $\alpha(z)$
couple to the physical degrees of the system, they can be neglected.

The coupled differential equations for this system nominally involves 
six degrees of freedom (one scalar and two gauge degrees of freedom for each
field). However, since the $x$ and $y$ components of each gauge
field enter quadratically into their respective Higgs field equations of 
motion, it is possible to rotate to a new basis $\tilde{x}$ and
$\tilde{y}$ where one only needs keep track of one component
of each gauge field. Note that the directions perpendicular to $z$ in which 
each of the gauge fields $A_1$ and $A_2$ point are independent. 
We therefore have only four degrees of freedom to non-trivially solve
for.

The equations we would like to solve are then
\begin{eqnarray}
A_1'' &=& 2e^2 R_1^2 A_1, \label{Gauge}\\
R_1'' &=& e^2A_1^2R_1 + 2\lambda_1 R_1(R_1^2+R_2^2-\upsilon^2)
+\lambda_2 R_1 R_2^2,
\label{Higgs}
\end{eqnarray}
and $1\leftrightarrow 2$. We have suppressed the spatial subscripts  
on the gauge fields, $A$.

For a domain wall solution the scalar fields must obey the boundary
conditions in eqn.~\ref{BC1}. Thus, by analyzing eqn.~\ref{Gauge}
we see that the gauge fields are required to have the following 
asymptotic behavior:
\begin{eqnarray}
A_1(z\rightarrow \infty) = e^{-\sqrt{2}e\upsilon|z|} \rightarrow 0
\quad {\rm and} \quad
A_2(z\rightarrow -\infty) = e^{-\sqrt{2}e\upsilon|z|} \rightarrow 0.
\label{A_assymp}
\end{eqnarray}
We observe that this asymptotic behavior is also consistent with 
eqn.~\ref{Higgs}.
The values of $A_1(-\infty)$ and $A_2(\infty)$ are seemingly unconstrained
by any of our differential equations. However, note that when 
$z\ll-1$ for $A_1(z)$ or when $z\gg 1$ for $A_2(z)$ the solutions 
become linear functions of $z$, the asymptotic solutions to eqn.~\ref{Gauge}. 
The linear solutions are due to the requirement that $R_1(z)$ and $R_2(z)$ 
vanish as $z\rightarrow - \infty, +\infty$ respectively (this is because we
require them to be kink solutions). Thus, the only allowed values of 
$A_1(-\infty)$ and $A_2(\infty)$ are either a {\it constant} (corresponding 
to constant asymptotic behaviour) or $\pm\infty$. 
Consistent with this, we will also impose the boundary conditions
\begin{equation}
A'_1(z = -\infty) = {\rm const.} \neq 0\quad {\rm and}\quad
A'_2(z =+\infty) = {\rm const.} \neq 0.
\label{Aprime_assymp}
\end{equation}
The requirement that these slopes be asymptotically nonzero removes the
$A_1 = A_2 = 0$ solution from our considerations.
Eqn.~\ref{Aprime_assymp} allows the constant slopes for 
$A_1$ and $A_2$ to be arbitrary. If they are chosen to be unequal, 
it implies 
that the corresponding magnetic fields $B_1$ and $B_2$ are unequal, 
leading to a violation of the symmetry inherent in the problem and 
this may also cause dynamical instability of the brane as will be 
discussed further in section 4. 
Hence, it is natural to choose the slopes to be equal. However, our 
numerical solutions (see Fig.~\ref{fig3}) show that even in the asymmetrical 
situation, slopes of $A_1$ and $A_2$ are very nearly equal.

The coupled differential equations \ref{Gauge} and \ref{Higgs}
together with
the conditions of eqns.~\ref{A_assymp} and \ref{Aprime_assymp}
constitute our boundary value problem (BVP). 

Since we shall resort to numerics to
find solutions it is convenient to transform from coordinate $z$ to
$u$ which is defined on a compact interval, $u\in[-1,1]$, via
\begin{eqnarray}
u=\tanh(\upsilon\sqrt{\lambda_1}z).
\end{eqnarray}
With this change of coordinates and the field rescalings
\begin{eqnarray}
R_a\rightarrow \upsilon R_a, \quad A_a \rightarrow \upsilon A_a, 
\end{eqnarray}
the equations become
\begin{eqnarray}
(1-u^2)^2{d^2 A_1\over du^2}-2u(1-u^2){d A_1\over du} &=& 2 \alpha R_1^2 A_1,
\label{eqone}\\
(1-u^2)^2{d^2 R_1\over du^2}-2u(1-u^2){d R_1\over du} &=& 
\alpha A_1^2 R_1 +2R_1(R_1^2+R_2^2-1) \nonumber\\
&& + \lambda R_1 R_2^2,
\label{eqtwo}
\end{eqnarray}
and $1\leftrightarrow2$. We have defined $\alpha\equiv e^2/\lambda_1$ and 
$\lambda\equiv\lambda_2/\lambda_1$. We see that solutions only 
depend on two independent coupling constants and not three.
In the case of the pure Higgs model
with $\alpha=0$ (see Ref.~\cite{clash}), if one takes symmetric ($R_1 + R_2$) 
and anti-symmetric ($R_1 - R_2$)
linear combinations of the fields, then the 
differential equations decouple for the special case of $\lambda=4$ 
with analytic solutions,
\begin{equation}
R_1={1\over 2}(1+u), \qquad R_2={1\over 2}(1-u).
\label{BC2}
\end{equation}
However, this is not the case in our model for $\alpha\neq0$.

We shall also be interested in the energy of the solutions we find, 
thus we need the stress-energy for this system
\begin{equation}
T_{\mu\nu} = 2{\delta {\cal L} \over \delta g^{\mu\nu}}-g_{\mu\nu}{\cal L},
\end{equation}
which for our action yields
\begin{eqnarray}
T_{\mu\nu} &=& -F_{1\mu\alpha} F^\alpha_{1\nu}
- F_{2\mu\alpha}F^\alpha_{2\nu}+2 (D_{1\mu}\phi_1)^* (D_{1\nu}\phi_1)
+ 2 (D_{2\mu}\phi_2)^* (D_{2\nu}\phi_2) \nonumber\\
&&+g_{\mu\nu}\biggl[
{1\over 4}F_1^{\mu\nu} F_{1\mu\nu}
+{1\over 4} F_2^{\mu\nu}F_{2\mu\nu}-(D_1^{\mu}\phi_1)^* (D_{1\mu}\phi_1)
-(D_2^{\mu}\phi_2)^* (D_{2\mu}\phi_2)  \nonumber\\
&&\hskip 1.3cm +V(\phi_1,\phi_2)\biggr].
\end{eqnarray}
The energy density is then given by the $T_{00}$ component of the
stress-energy tensor. This simplifies to 
\begin{eqnarray}
T_{00} &=& 
{1\over 4}\left[\left(A'_1(z)\right)^2
+\left(A'_2(z)\right)^2\right]
+\left(R'_1(z)\right)^2
+\left(R'_2(z)\right)^2 \nonumber\\
&&  +e^2 A_1(z)^2 R_1(z)^2
+e^2 A_2(z)^2 R_2(z)^2 +V(R_1,R_2),
\end{eqnarray}
for our static solutions and because of our gauge choice, $A_0=0$. 
Thus, in terms of the coordinate $u$ and the rescaled fields the
energy density is given by
\begin{eqnarray}
{T_{00}\over \lambda_1\upsilon^4} &=& 
(1-u^2)^2\left[
{\left(\partial_u A_1(u)\right)^2\over 4}
+{\left(\partial_u A_2(u)\right)^2\over 4}
+\left(\partial_u R_1(u)\right)^2
+\left(\partial_u R_2(u)\right)^2 \right]\nonumber\\
&&  +\alpha A_1(u)^2 R_1(u)^2
+\alpha A_2(u)^2 R_2(u)^2 +\left(R_1(u)^2+R_2(u)^2-1\right)^2 \nonumber\\
&&+\lambda R_1(u)^2 R_2(u)^2,
\end{eqnarray}
where $\lambda_1\upsilon^4$ sets the scale.

\section{Numerical Solutions}
  
The numerical method we employ to solve these coupled differential equations
is the `shooting method' using the routines from Numerical Recipes in
C++~\cite{NumericalRecipes}. One can readily 
convert our system of four coupled second order
differential equations to a system of eight coupled first order
differential equations where the functions are: $R_1$, $R_2$, $A_1$, $A_2$,
$R_1'$, $R_2'$, $A_1'$ and $A_2'$. This is a boundary value problem 
with the functions $R_1$, $R_2$, $A_1$, $A_2$ are specified on
two boundaries but with the functions $R_1'$, $R_2'$, $A_1'$, $A_2'$ not specified
on either boundary. The way the `shooting method' works is that one
guesses values for the derivative functions at the left boundary
($u=-1$), then with all the functions specified on the left boundary
one can numerically integrate to the right boundary. One then defines 
a function which measures how well the boundary conditions on the right
are matched. Using this goodness of fit function one can then use 
a Newton-Raphson procedure to improve the guess 
on the left boundary for the derivatives.
One can then iterate this procedure until the boundary conditions on
both sides are satisfied to the desired accuracy. One potential
difficulty is that if the
differential equations are reasonably complicated (e.g. non-linear)
then the {\it initial guess} might need to be reasonably good in order
for the procedure to converge.

The differential equations, \ref{eqone} and \ref{eqtwo},  
have poles at $u=\pm1$ when one
expresses the equations as $d X/du = (1-u^2)^{-2}\times\ldots$ 
Since we cannot evaluate these equations at $u=\pm1$, we set the boundaries
at $u_1=-1+\epsilon$ and $u_2=1-\epsilon$. However, because
now our 
boundaries are not at $u=\pm1$ ($z=\pm\infty$) we need to know the
asymptotic behavior of our functions in order to set up the 
boundary conditions correctly\footnote{For numerical reasons 
we can not just set $R_1(u_1)=0$, $R_1(u_2)=1$ \ldots}. 
For the special case of 
$\alpha=0$ and $\lambda=4$ the analytic solution, eqn.~\ref{BC2}, 
is known from Ref.~\cite{clash}. 
While these are not the correct solutions for general $\alpha$ and 
$\lambda$,  they do exhibit the correct asymptotic behavior as 
$u\rightarrow\pm1$. But as long as $\epsilon$ is sufficiently small 
the correct asymptotic behavior is obtained numerically.
When we solve our boundary value problem numerically 
we shall use eqn.~\ref{BC2} to set the boundary conditions for $R_1$ 
and $R_2$. We also need to know the asymptotic behavior of the gauge
fields near the boundaries. Substituting $A=(1-u^2)^\beta$ 
into the differential equation for $A$ (eqn.~\ref{eqone}), 
we can solve for $\beta$,
the scaling behavior in the vicinity of the boundary. Thus
\begin{eqnarray}
A_1  \sim (1-u)^{\sqrt{\alpha/2}} \sim \epsilon^{\sqrt{\alpha/2}}
& \quad {\rm as} \quad & u\rightarrow u_2=1-\epsilon \\
A_2  \sim (1+u)^{\sqrt{\alpha/2}} \sim \epsilon^{\sqrt{\alpha/2}}
& \quad {\rm as} \quad & u\rightarrow u_1=-1+\epsilon.
\end{eqnarray}
The values of $A_1(u_1)$ and $A_2(u_2)$ are not constrained by
any of the differential equations and are therefore left as
free parameters.

As mentioned before when solving a boundary value problem using 
the `shooting method', convergence may depend on a reasonably accurate
guess of the initial conditions on the left boundary. This is the case
for our set of differential equations since they have an explicit
pole at $u=\pm 1$. This sensitivity gets worse as $\epsilon$ approaches
zero. The method we employed to address this issue involved starting 
with a relatively large value of $\epsilon$ ($\epsilon=0.5$) and
incrementally reducing it to its desired value using as the 
initial guess for the values of the derivatives ($R_1'$, $R_2'$, $A_1'$ 
and $A_2'$) on the left boundary for each step the solution of 
the previous step.

\begin{figure}[ht]
\centerline{\epsfig{file=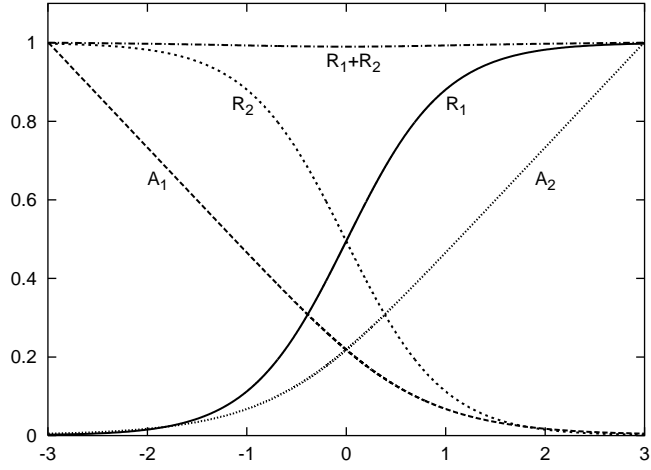,width=3.5in}}
\caption[]{Plot of $R_1$, $A_1$, $R_2$, $A_2$ and $R_1+R_2$ 
against $\tanh^{-1}(u)$ for $\alpha=1$, $\lambda=4$. The 
free boundary conditions are $A_1(-1+\epsilon)=A_2(1-\epsilon)=1$
for $\epsilon=0.005$ which corresponds to left and right boundaries 
at $\tanh^{-1}(u)=\pm3$. $R_1+R_2$ is nearly constant for this 
pair of parameters.}
\label{fig1}
\end{figure}
\begin{figure}[ht]
\centerline{\epsfig{file=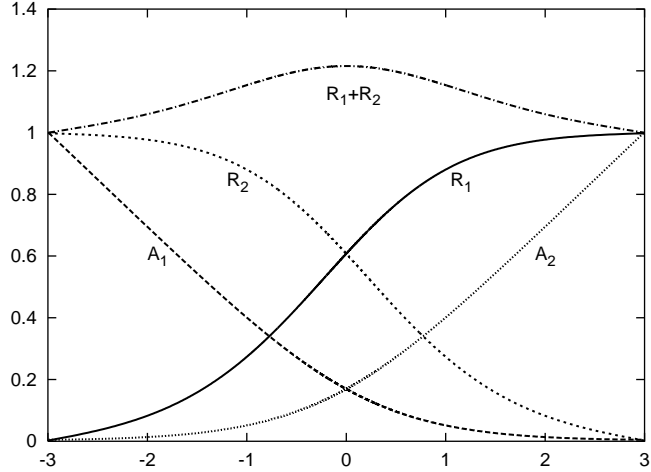,width=3.5in}}
\caption[]{Plot of $R_1$, $A_1$, $R_2$, $A_2$ and $R_1+R_2$ 
against $\tanh^{-1}(u)$ for $\alpha=1$, $\lambda=1$. Here 
$A_1(-1+\epsilon)=A_2(1-\epsilon)=1$ for $\epsilon=0.005$.}
\label{fig2}
\end{figure}
\begin{figure}[ht]
\centerline{\epsfig{file=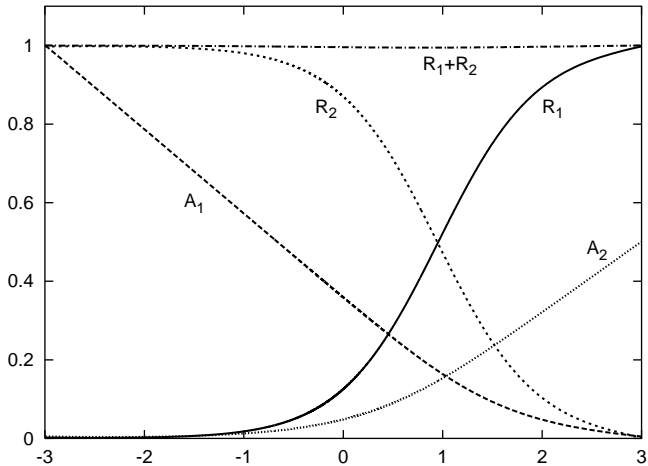,width=3.5in}}
\caption[]{Plot of $R_1$, $A_1$, $R_2$, $A_2$ and $R_1+R_2$ 
against $\tanh^{-1}(u)$ for $\alpha=1$, $\lambda=4$. Here 
$A_1(-1+\epsilon)=1$ and $A_2(1-\epsilon)=0.5$ for $\epsilon=0.005$. 
The principal effect of the asymmetrical BC's is to shift the center of the 
brane to the right. }
\label{fig3}
\end{figure}
In figs.~\ref{fig1}, \ref{fig2} and \ref{fig3}  we see numerical
solutions to these differential equations for a variety of 
couplings, $\alpha$, $\lambda$ and boundary conditions $A_1(u_1)$ 
and $A_2(u_2)$. We observe that the gauge fields $A_1$ and $A_2$
become linear functions of $\tanh^{-1}(u)$ as $u\rightarrow u_1$ and
$u\rightarrow u_2$ respectively. This implies that asymptotically
these gauge fields become linear functions of $z$, which corresponds
to a constant magnetic field in the direction perpendicular to both
$z$ and $\tilde{x}$ (the direction in which the gauge field points),
\begin{equation}
B_{\tilde{y}} \sim \partial_z A_{\tilde{x}}(z) = {\rm constant}.
\end{equation}
Thus the asymptotic solution (actually $\tanh(u)^{-1}$ need only be
of the order of $\pm 2$ to be in the asymptotic regime for a 
typical configuration) on either side of the domain wall is
a constant magnetic field corresponding to the $U(1)$ fields, which point 
in uncorrelated directions parallel to the domain wall. These 
solutions have non-zero energy density away from the domain wall 
and thus are infinite energy configurations. The solutions where
the magnetic fields are both zero corresponds to the choice of
$\alpha=0$ (i.e. no $U(1)$ gauge fields). 

In figs.~\ref{fig1} and \ref{fig2}, we have set 
$A_1(u_1=-0.995)=A_2(u_2=0.995)=1$. With this set of symmetric
boundary conditions the domain wall is centered at $u=0$. In 
fig.~\ref{fig3} we see that the effect of asymmetric BC's is to 
shift the location of the domain wall. While not apparent in the 
figure the magnitudes of the uniform magnetic field far from either
side of the domain wall do not exactly match.
The choice of $\epsilon=0.005$ (and $\epsilon=0.001$ for figs.~\ref{energy} 
and \ref{zeropoint}) corresponds to boundaries at 
$\tanh(u)^{-1}=\pm 3$ (and $\pm 3.8$). 
While $\epsilon$ can be made 
smaller at the expense of longer computing time, these values are 
sufficiently small for our purposes.

\begin{figure}[ht]
\centerline{\epsfig{file=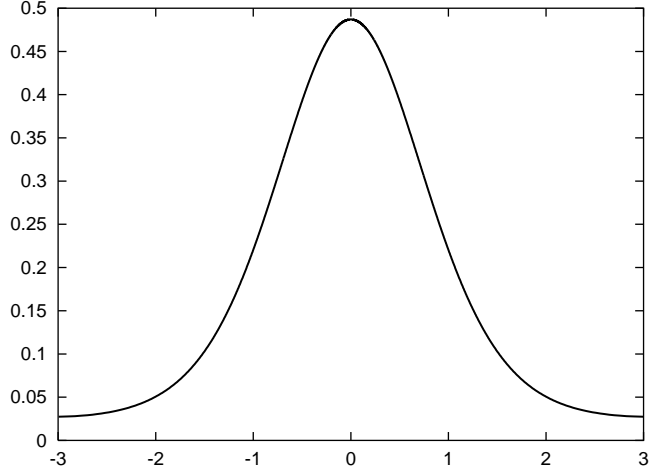,width=3.5in}}
\caption[]{Plot of the energy density against $\tanh^{-1}(u)$ for 
$\alpha=1$, $\lambda=1$. We have used the boundary conditions 
$A_1(-1+\epsilon)=A_2(1-\epsilon)=1$ where $\epsilon=0.005$.}
\label{energy_den}
\end{figure}
\begin{figure}[ht]
\centerline{\epsfig{file=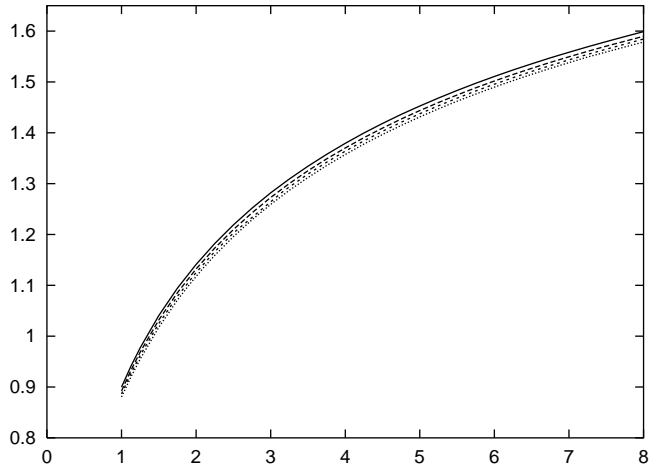,width=3.5in}}
\caption[]{Plot of the `renormalized' surface energy density against 
$\lambda$ for $\alpha=0.25$, 
$0.5$, $1.0$, $2.0$ (from top to bottom). We have used the boundary 
conditions $A_1(-1+\epsilon)=A_2(1-\epsilon)=1$ where  $\epsilon=0.001$.}
\label{energy}
\end{figure}
\begin{figure}[ht]
\centerline{\epsfig{file=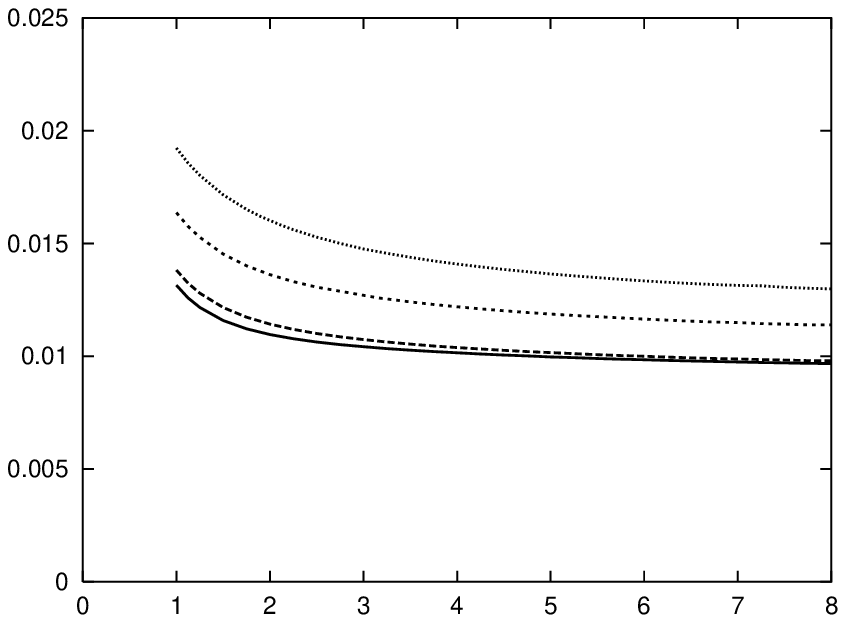,width=3.5in}}
\caption[]{Plot of the energy density of the uniform magnetic field 
against $\lambda$ for 
$\alpha=0.25$, $0.5$, $1.0$, $2.0$ 
(from bottom to top). We have used the boundary conditions 
$A_1(-1+\epsilon)=A_2(1-\epsilon)=1$ where $\epsilon=0.001$.}
\label{zeropoint}
\end{figure}
In fig.~\ref{energy_den} we see the energy density of a solution 
plotted as a function of the transverse direction. We 
see that the energy density is peaked at the center of the
domain wall. If we treat the asymptotic constant magnetic field on either
side of the domain wall as a background, then we can compute 
the energy per unit surface area of the domain wall by subtracting
off the infinite energy associated with the magnetic field. 
In fig.~\ref{energy} the surface energy density is plotted
as a function of $\lambda$ for a variety of values of $\alpha$ 
(boundary conditions are $A_1(-1+\epsilon)=A_2(1-\epsilon)=1$ at 
$\epsilon=0.001$). Observe that this `renormalized'
surface energy density is only
weakly dependent on the value of the gauge coupling constant $\alpha$.
In fig.~\ref{zeropoint} we show the subtracted
energy density corresponding to the constant magnetic field as
a function of $\lambda$ and $\alpha$. In both figs.~\ref{energy} and
\ref{zeropoint} we omit values of $\lambda<1$ as they require 
a significantly smaller value for $\epsilon$.

Our solutions are all plotted in units of $\tanh(u)^{-1}$ and not $z$
since $u$ is the natural variable in our system of equations, \ref{eqone} and 
\ref{eqtwo}. The length scale $\tanh(u)^{-1}$ is dimensionless and can be
converted into a physical length by dividing by $\upsilon\sqrt{\lambda_1}$.
The thickness of the domain wall is typically 
$\sim 4 / \upsilon\sqrt{\lambda_1}$ (see fig.~\ref{energy_den}) which
can be made arbitrarily small by choosing $\upsilon\sqrt{\lambda_1}\gg 1$.

\section{Discussion}

The numerical solutions displayed above have a natural interpretation
in terms of superconductivity. Consider, for instance, the currents associated
with the $U(1)$ gauge groups,
\begin{equation}
J_{i\,\mu} = i e [ \phi^*_i (\partial_{\mu} \phi_i) - (\partial_{\mu} \phi_i^*)
\phi_i] + 2 e^2 A_{i\,\mu} \phi_i^* \phi_i,
\end{equation}
where $i=1,2$. In terms of the amplitude and phase of $\phi_i$, the currents
are given by
\begin{equation}
J_{i\,\mu} = -2e R^2_i \partial_{\mu} \alpha_i + 2 e^2 A_{i\,\mu} R_i^2.
\end{equation}
For our configurations, which depend only on $z$, and for which eqn.~\ref{PureGauge}
holds, it is clear that only the $x$- and $y$-components are non-vanishing.
They evaluate to
\begin{equation}
J_{i\,x,y}(z) = 2 e^2 A_{i\, x,y}(z) R_i^2(z).
\label{Jresult}
\end{equation}
These steady, $z$-dependent current densities are
uniform supercurrent densities localised to the domain wall, with the
charged boson fields as the current carriers. 

Equation~\ref{Jresult} shows
that the currents are nonzero only when the gauge field configurations are
nonzero and vice-versa, so these currents are responsible for
dynamically generating the magnetic fields. On the side of the wall where
$R_i \neq 0$, the corresponding magnetic field is seen to decay exponentially, 
which is simply a Meissner effect. On the other side of the wall, where $R_i$
is tending exponentially quickly to zero, we find the magnetic
field $\vec{B}_i$ tending towards a finite, 
uniform configuration pointing in the plane of the wall.
This is consistent with the domain wall carrying a uniform sheet of
current density pointing in the $(0,A_{i\,x},A_{i\,y},0)$
direction, as per eqn.~\ref{Jresult}. Our configurations have infinite
energy because the domain wall is of infinite extent, with current
densities uniformly distributed on it. 

The stability or otherwise of our solutions is an important concern. While a
complete stability analysis is beyond the scope of this paper, the
above considerations suggest that the geometrically symmetric solutions
such as in figs.~\ref{fig1} and \ref{fig2} could be stable, whereas asymmetric
configurations such as those of fig.~\ref{fig3} are not. Let current
$J_1$ point in the $x$-direction in the plane of the wall. Then eqn.~\ref{Jresult}
implies that $A_1$ also points in the same direction, so $\vec{B}_1$
is directed along the $y$-axis. The Lorentz force on the type 1 charge carriers
lies in the negative $z$ direction. For sector 2, similar reasoning
shows that the corresponding Lorentz force on type 2 charge carriers points
in the positive $z$ direction. For symmetric boundary conditions, these
forces are equal in magnitude as well as opposite in direction. This is
a necessary condition for stability. 
For asymmetric boundary conditions,
they are unequal, strongly suggesting that such configurations are unstable.


\section{Conclusions}

In order to further explore the idea of the ``clash of symmetries'' 
from \cite{clash}, we have considered a model in which two scalar 
fields are coupled to their respective gauge fields in a
Lagrangian which has $U(1)\times U(1)$ symmetry. We find 
consistent static solutions for field configurations with the 
vacuum conditions for the scalar fields specified by eqn.~\ref{BC1} and 
the implied boundary conditions for the gauge fields, eqn.~\ref{A_assymp}. 
We obtain the expected kink-like solutions for the scalar fields 
while the two gauge fields diverge linearly on either side of the 
domain wall.

When we consider the idealized configuration of an infinitely thin 
domain wall, we have solutions such that while the $U(1)$ symmetries 
of the fields are preserved in their respective vacua, they are 
both broken on the domain wall. The gauge fields show that the 
domain wall is sandwiched between domains with constant magnetic 
fields parallel to the wall. In the case of a domain wall of finite
thickness, there will be magnetic fields parallel to the wall on
either side. These are associated with superconducting currents, as in 
the case of the superconducting string solution~\cite{Witten}.

This model demonstrates that in addition to the breakdown of
symmetries on the brane, the presence of gauge fields introduces
new phenomena, such as the appearance of magnetic fields. 
Background magnetic fields of this kind are reminiscent of the 
configurations in string theory that give rise to non-commutativity 
of space-time coordinates. It would be very interesting to see the 
logical extension of this model to domain wall solutions with 
non-Abelian gauge fields and to study their dynamical effects in 
addition to symmetry breaking.

\acknowledgments

This work was supported in part (JSR and KCW) by the U.S Department 
of Energy (DOE) under contract no. DE-FG02-85ER40237. 
RRV was supported partially by the Australian Research Council
and partially by the University of Melbourne. He thanks his
coauthors for their excellent hospitality at Syracuse University
where this work was begun.


\begin{thebibliography}{1}

\bibitem{Witten}
E.~Witten,
Nucl.\ Phys.\ B {\bf 249}, 557 (1985).

\bibitem{MacKenzie}
R.~MacKenzie,
Nucl.\ Phys.\ B {\bf 303}, 149 (1988).

\bibitem{Lemperiere}
Y.~Lemperiere and E.~P.~Shellard,
Nucl.\ Phys.\ B {\bf 649}, 511 (2003)
[arXiv:hep-ph/0207199].

\bibitem{clash}
A.~Davidson, B.~F.~Toner, R.~R.~Volkas and K.~C.~Wali,
Phys.\ Rev.\ D {\bf 65}, 125013 (2002)
[arXiv:hep-th/0202042].

\bibitem{pv}
L.~Pogosian and T.~Vachaspati,
Phys.\ Rev.\ D {\bf 62}, 123506 (2000)
[arXiv:hep-ph/0007045];\\
see also T.~Vachaspati,
Phys.\ Rev.\ D {\bf 63}, 105010 (2001)
[arXiv:hep-th/0102047];\\
L.~Pogosian and T.~Vachaspati,
Phys.\ Rev.\ D {\bf 64}, 105023 (2001)
[arXiv:hep-th/0105128].

\bibitem{Shin}
E.~Shin and R.~R.~Volkas, Phys.\ Rev.\ D (in press)
[arXiv:hep-ph/0309008].

\bibitem{Morris}
J.~R.~Morris,
Phys.\ Rev.\ D {\bf 52}, 1096 (1995).

\bibitem{Lazarides}
G.~Lazarides and Q.~Shafi,
Phys.\ Lett.\ B {\bf 159}, 261 (1985).

\bibitem{NumericalRecipes}
B.~P.~Flannery, W.~H.~Press, S.~A.~Teukolsky and W.~T.~Vetterling,
``Numerical Recipes in C++,'' 
(second edition), Cambridge University Press, New York, 2002.

\end{thebibliography}
\end{document}